\documentclass[aps, prl, superscriptaddress, twocolumn]{revtex4}

\usepackage{amssymb,amsmath}
\usepackage{graphicx}
\usepackage{verbatim}
\usepackage{color}
\usepackage{subfigure}
\usepackage{float}

\begin{document}
\title{A density matrix approach for the electroluminescence of molecules
in a scanning tunneling microscope}
\date{\today}

\author{Guangjun Tian}
\affiliation{Hefei National Laboratory for Physical Sciences at
the Microscale, University of Science and Technology of China,
Hefei, Anhui 230026}
\affiliation{Theoretical Chemistry, School of Biotechnology,
Royal Institute of Technology, S-106 91 Stockholm, Sweden}
\author{Jicai Liu}
\affiliation{Theoretical Chemistry, School of Biotechnology,
Royal Institute of Technology, S-106 91 Stockholm, Sweden}
\author{Yi Luo}
\email[e-mail: ]{luo@kth.se}
\affiliation{Hefei National Laboratory for Physical Sciences at
the Microscale, University of Science and Technology of China,
Hefei, Anhui 230026}
\affiliation{Theoretical Chemistry, School of Biotechnology,
Royal Institute of Technology, S-106 91 Stockholm, Sweden}

\begin{abstract}
The electroluminescence of molecules confined inside a nanocavity in
the scanning tunneling microscopy possesses many intriguing
but unexplained features. We present here a general theoretical
approach based on the density matrix formalism to describe the
electroluminescence from molecules near a metal surface induced by
both electron tunneling and local surface plasmon excitations simultaneously.
It reveals the underlying physical mechanism for the external bias dependent
electroluminescence. The important role played by the local surface
plasmon on the electroluminescence is highlighted. Calculations for
porphyrin derivatives have reproduced corresponding experimental
spectra and nicely explained the observed unusual
large variation of emission spectral profiles. This general theoretical
approach can find many applications in the design of molecular electronic
and photonic devices.
\end{abstract}

\maketitle
Light emission of molecules induced by the localized tunneling current
in a scanning tunneling microscope (STM) has been studied for almost two decades
\cite{Johansson_PRB_1990, Berndt_PRL_1991, Berndt_Science_1993, Johansson_PRB_1998,
Aizpurua_PRB_2000, qiu_science_2003, dong_prl_2004, Cavar_PRL_2005,
buker_prb_2005, Ho_PRB_2008, buker_prb_2008, seldenthuis_prb_2010,
Rossel_SSR_2010}, owing to its rich new physics and fascinating potential
applications in molecular scale optics and electronics\cite{qiu_science_2003,dong_prl_2004,Cavar_PRL_2005,Ho_PRB_2008,dong_natphotonics_2009}.
In principle, the electroluminescence of a molecule can be generated once an electron-hole pair
is formed inside the molecule. This could happen when the highest occupied molecular orbital
(HOMO) and the lowest unoccupied molecular orbital (LUMO) enter into the
window of the external bias\cite{qiu_science_2003,dong_prl_2004}. Such an electron
driven mechanism has been explained by several groups using different
theoretical methods\cite{buker_prb_2005, buker_prb_2008, seldenthuis_prb_2010}.
Very recently, a fascinating new experimental study by Dong \emph{et al.}\cite{dong_natphotonics_2009}
has shown that the electroluminescence spectra of tetraphenyl porphyrin (TPP) molecules
can be effectively modified by tuning the energy profile of the resonant nanocavity plasmons (NCPs)
in a STM. The resonant hot electroluminescence arising directly from higher vibronic
levels of the singlet excited state for porphyrin molecules, even upconversion luminescence,
have been observed for the first time. It was suggested that the local nanocavity plasmons
behave like a strong coherent optical source with a tunable energy\cite{dong_natphotonics_2009}.
However, there is no fully developed theoretical model ready to describe these new phenomenons.

It was known that the light induced fluorescence and Raman scattering of molecules in the
metallic nano-gap could be uniformly treated by the density matrix approach\cite{xu-prl,jonansson-prb}.
In this letter, we will show that the density matrix formulation can be generalized
to describe the electroluminescence from molecules near a metal surface in a STM.
In our model, the surface plasmons are regarded as strong coherent electromagnetic sources,
the electron tunneling rates are determined by Fermi's golden rule, and the finite lifetimes
of the energy levels are introduced phenomenologically into the density matrix equations
as relaxations. In other words, the spontaneous, electron tunneling induced, and stimulated
emissions can be uniformly described by this newly proposed approach.

The equation of motion of the density matrix is:
\begin{equation}
      \frac{\partial \rho}{\partial t}=-\frac{\imath }{\hbar}
      [\hat{H}, \rho ]+\Gamma_{tr}\rho+\Gamma_{ph}\rho+R\rho
\label{eq:denmat}
\end{equation}
Here $\rho$ is the Hermitian density operator, its diagonal matrix element
$\rho_{nn}$ gives the probability of the molecule in the level $n$.
Within the electric dipole approximation, the total Hamiltonian of the system
$\hat{H}$ is the sum of the free system Hamiltonian $\hat{H_{0}}$ and the
interaction Hamiltonian of the system and the plasmon,
$\hat{V}(\boldsymbol{r}, t)=-\boldsymbol{\mu \cdot\varepsilon}$.
Considering the surface plasmon as a linearly polarized electromagnetic field,
we have $\boldsymbol{\varepsilon}(t)= \mathbf{e}E_{0} cos(\omega t)$.
Here $\mathbf{e}$ and $\omega$ are the polarization vector and the frequency of
the surface plasmon, respectively. $\boldsymbol{\mu}$ is the electric dipole moment operator
of the molecule. $\Gamma_{tr}$ is the population decay operator with\cite{xu-prl},
\begin{equation}
\Gamma_{tr}\rho=-\sum_{ij}\frac{\Gamma_{i\leftarrow j}}{2}[a_{ji}a_{ij}\rho
-2a_{ij}\rho a_{ji}+\rho a_{ji} a_{ij}]
\end{equation}
Here $\Gamma_{i\leftarrow j}$ is the decay rate from the level $j$ to the level $i$.
$a_{ij}$ is a matrix with element $(i, j)$ equals to 1 and all other elements
equal to 0. $\Gamma_{ph}$ is the dephasing operator which is introduced
into the density matrix to describe the broadening of the emission spectrum
caused by the surroundings.

The electron tunneling process can only affect the population of the molecule
through the diagonal terms of the density matrix (tunnel into and out of a
particular level). The last term in Eq. ~(\ref{eq:denmat}) has the form,
\begin{equation}
R\rho = -\sum_{ij}[R_{j\to i}a_{jj}\rho a_{jj}-R_{i\to j}a_{ji}\rho a_{ij}]
\end{equation}
Here $R_{i \to j}$ ($R_{j \to i}$) is the electron tunneling rate from level $i$ to $j$
($j$ to $i$). Under the approximation of weak coupling between the molecule and the
STM tip and substrate, these rates can be calculated according to Fermi's golden rule as
widely used in the rate equation method\cite{bonet_prb_2002, seldenthuis_acsnano_2008,seldenthuis_prb_2010}.
If we assume the STM tip as the left lead and the substrate as the right lead to the molecule.
The tunneling rates between two relevant levels $i$ and $j$ ($E_{i}<E_{j}$) could be obtained
\cite{mccarthy-prb-2003, koch-prb-2006, seldenthuis_acsnano_2008},
\begin{align}
R_{i\rightarrow j}&=F_{ij}[\gamma^{L}f(\delta E^{L})+\gamma^{R}f(\delta E^{R})]\nonumber \\
R_{j\rightarrow i}&=F_{ji}[\gamma^{L}(1-f(\delta E^{L}))+\gamma^{R}(1-f(\delta E^{R}))]
\end{align}
with $\delta E^{L}=E_{j}-E_{i}-e\alpha V$ and $\delta E^{R}=E_{j}-E_{i}-e(\alpha-1)V$.
$F_{ij}$ is the transition probability between the levels $i$ and $j$.
$\gamma^{R}$ and $\gamma^{L}$ are the bare tunneling rates between a particular
level of the molecule and the left and right leads (for the simplicity,
we assume different levels have the same bare tunneling rates), $\alpha$ is
the bias coupling over the leads, $f$ is the Fermi distribution at the
temperature $T$, $f(E)=1/[1+\mathrm{exp}(E/k_{B}T)]$.

Relative to the level $i$ in the ground state, the probability of
radiative population or depopulation of the level $j$ in the excited
state induced by the plasmon is depicted by the work of the field $W_{ji}^{p}(t)$,
\begin{equation}
W_{ji}^{p}(t)=\text{Im}[H_{ji}\rho_{ij}-\rho_{ji}H_{ij}]
\end{equation}
Determined by the instantaneous Rabi frequency, $W_{ji}^{p}(t)$ shows
a sign-changing property with its positive ($W_{ji}^{pA}(t)$) and negative
($W_{ji}^{pE}(t)$) parts representing respectively the absorption and the
stimulated emission induced by the plasmon. Correspondingly, the work done
by the electron tunneling process at time $t$ can be calculated as,
\begin{equation}
W_{ji}^{e}(t)=W_{ji}^{eA}(t)+W_{ji}^{eE}(t)=R_{i\to j}\rho_{ii}-R_{j\to i}\rho_{jj}
\end{equation}
The contributions of spontaneous emissions to the total emission spectra
are described by the radiative decay from level $j$ in the excited
state to the level $i$ in the ground state,
$W_{ji}^{sE}(t)=-\Gamma_{i\leftarrow j}\rho_{jj}$.
All of these quantities can be calculated by solving the density matrix
equation of the system, and the emission cross section for the transition
between the level $j$ and the level $i$ can then be obtained from,
\begin{equation}
\sigma _{ji}=A \int_{-\infty }^{\infty }(W_{ji}^{pE}(t)+W_{ji}^{eE}(t)+W_{ji}^{sE}(t))dt
\label{eq:intensity}
\end{equation}
$A$ at the right-hand side of the equation is a constant prefactor.
The photon emission process can be described by a Lorentzian function,
the power spectrum is then obtained as,
\begin{equation}
\sigma(\omega )= \frac{1}{\pi} \sum_{j}\sum_{i}
\frac{\sigma_{ji}\tau }{(\omega _{ji}-\omega)^2+\tau ^2}
\label{eq:conv}
\end{equation}
Here $\tau$ is the half width at half maximum (HWHM) of the emission line,
$\omega _{ji}$ is the resonant frequency between the levels $j$ and $i$.
It is noted that the system can reach its steady state in several nanoseconds for
all the cases studied here, which is several orders of magnitude
shorter than the time of the experimental measurements. Thus the
contribution from the transient states can be averaged out.

\begin{figure}[ht]
\begin{center}
\includegraphics[width=8.2cm]{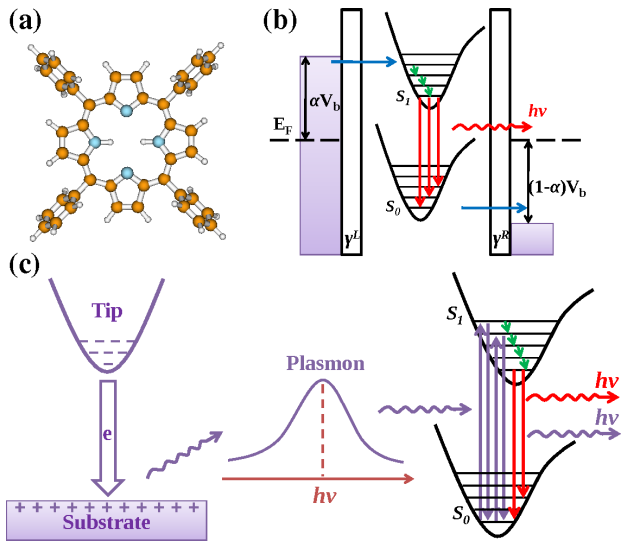}
\caption{\label{fig:fig1} (a) The optimized structure of a tetraphenyl porphyrin (TPP)
molecule. (b) Electron tunneling induced emission with the inclusion of the internal
relaxation. (c) Plasmon-assist process, including the
plasmon generation, the excitation, the stimulated emission and
the spontaneous emission.}
\end{center}
\end{figure}

We have solved the related density matrix equations for a TPP molecule
(optimized structure shown in Fig.\ref{fig:fig1}(a)) in a STM as done in
the latest experiment of Dong \emph{et al.}\cite{dong_natphotonics_2009}.
The three processes involved in the electroluminescence from the molecule
are schematically shown in Fig.\ref{fig:fig1}. The electron
tunneling induced emission process is illustrated in Fig.\ref{fig:fig1}
(b), in which electrons can tunnel into the LUMO and tunnel out from the HOMO
when both orbitals are in the bias window and leave an electronic excited state
ready to fluoresce. Due to the relative short lifetime of the vibrational
excited state, the system will first relax to the lowest vibrational state
of the excited state ($S_1$) before emitting a photon according to the Kasha's
rule. In the plasmon-assist molecular emission process (Fig.\ref{fig:fig1}
(c)), a plasmon can be generated by the inelastic tunneling of electrons from
the tip to the metal substrate (IET process). The plasmon behaves as a strong
coherent electromagnetic source and can excite the molecule resonantly into
higher vibrational levels in the first excited state following a resonant
emission, or stimulated emission, without a non-radiative relaxation\cite{dong_natphotonics_2009}.
It is noted that when the plasmonic excitation is relative weak, the emission
will again follow the Kasha's rule. Spontaneous emission (fluorescence) has a
similar mechanism with the electron tunneling induced emission and
it happens simultaneously with the other two emission processes.

In principle, our formulation allows to include as many electronic states and
vibrational levels as one wants to. However, for the relevant emission
processes of the TPP molecule, it is enough to consider two electronic states,
$S_{0}$ and $S_{1}$ (Q band), each with three vibrational levels.
The transition probability between different vibrational levels
is governed by the Franck-Condon (FC) factors that can be calculated using the
standard harmonic model. The calculated energy gap between the two vibrational
bands, $Q(0, 0)$ and $Q(0, 1)$, is scaled to the experimental value of
0.16eV and the excitation energy between two molecular states is set
to the experimental value of 1.89 eV\cite{dong_natphotonics_2009} for the sake of
a better presentation. The initial condition for the population is
$\rho_{0, 0}=1$ and $\rho_{n, n}=0, n\neq0$, i.e. the electrons rest at the
lowest vibrational level of the ground state. Normally,
the radiative lifetime of the state is about three orders of magnitude longer than
those of non-radiative relaxations\cite{gelmukhanov_josab_2002}. In this work the radiative
lifetime $\tau_{r}$ is set to be 2 ns and the radiative decay rate between level $j$ in $S_{1}$
and $i$ in $S_{0}$ is calculated as $\Gamma _{i\leftarrow j}=F_{ji}/\tau_{r}$. The
non-radiative decay rates between the vibrational levels are set to be (2 ps)$^{-1}$.
The density matrix equations are solved using an iterative predictor-corrector method.
The dephasing factor enters the calculated spectrum through 
the HWHM ($\tau=0.05$ eV for all the calculations in this work)
of the convoluted Lorentzian profile (Eq.~(\ref{eq:conv})).
All the calculations are performed with a bias coupling constant
($\alpha$) of 0.50 and nearly symmetric bare tunneling rates ($\gamma^{L}$=16.40$\mu \textup{eV}$,
$\gamma^{R}$=4.80$\mu \textup{eV}$) as those used by Seldenthuis \emph{et al.}\cite{seldenthuis_prb_2010}
for a similar system.

\begin{figure}[ht]
\begin{center}
\includegraphics[width=8.4cm]{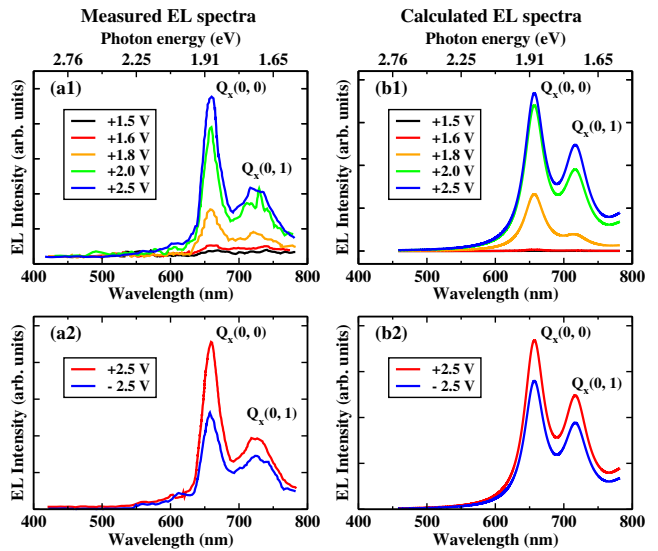}
\caption{\label{fig:fig2}(a1) and (a2) The measured EL spectra of H$_{2}$TBPP molecules
at different bias voltages. [Adapted with permission from Dong \emph{et al.} (Ref.
\onlinecite{dong_prl_2004}) copyright (2004) by the American Physical Society.]
(b1) and (b2) The calculated EL spectra of TPP molecules at different bias voltages
without considering the plasmon excitation. The temperature is set to be 300 K.}
\end{center}
\end{figure}

We first simulate the electroluminescence (EL) spectra of a TPP molecule in a STM
induced by the electron tunneling process only. The bias voltage dependence of
EL spectra for different porphyrin molecules were experimentally
observed by Qiu \emph{et al.}\cite{qiu_science_2003} and Dong
\emph{et al.}\cite{dong_prl_2004}. It was found that the spectral peaks
under different bias voltages remain unchanged with enhanced intensities
at higher bias voltages. Dong \emph{et al.} also reported bipolar EL spectra
at $\pm$ 2.5V. In a very recent theoretical study by Seldenthuis \emph{et al.}
\cite{seldenthuis_prb_2010}, the bias voltage dependence of EL spectra
measured by Qiu \emph{et al.} as well as the bipolar molecular fluorescence
observed by Dong \emph{et al.} have been successfully analyzed using a rate equation method.
We focus our attention to the results for the \textit{meso}-tetrakis
(3,5-di-terbutylphenyl) porphyrin (H$_{2}$TBPP) molecules\cite{dong_prl_2004}.
Our simulated spectra under different bias voltages from the density matrix approach
are illustrated in Fig.~\ref{fig:fig2}(b1), which have largely reproduced the
corresponding experiments as recaptured in Fig~\ref{fig:fig2}(a1).
The peak positions of the calculated EL spectra all remain the same simply due to
the Kasha's rule. A clear enhancement of the spectral intensities at higher bias
voltages is revealed, which is arisen from the increased number of vibrionic levels
entering in the bias window. One interesting experimental observation is that the EL can still
be observed when the excitation voltage is less than the energy of the emitted photon
around 1.9 eV. Such "energy forbidden" transitions were suggested to be resulted from
either the thermally assisted tunneling injection of holes or the shift of energy
levels due to the charging of the molecules\cite{dong_prl_2004}. Our simulations
have shown that the thermally assisted electron tunneling is the reason because of the broad
Fermi-Dirac distribution at 300K. Moreover, the experimentally observed bipolar spectra
at $\pm$ 2.50 V in Fig.~\ref{fig:fig2}(a2) are also well reproduced by our
calculations (Fig.~\ref{fig:fig2}(b2)) as a result of the weak couplings
between the leads and the molecules. The intensity difference
in the bipolar spectra is due to the slight asymmetric coupling between the molecule
and the two leads\cite{seldenthuis_prb_2010}.

\begin{figure}[ht]
\begin{center}
\includegraphics[width=8.6cm]{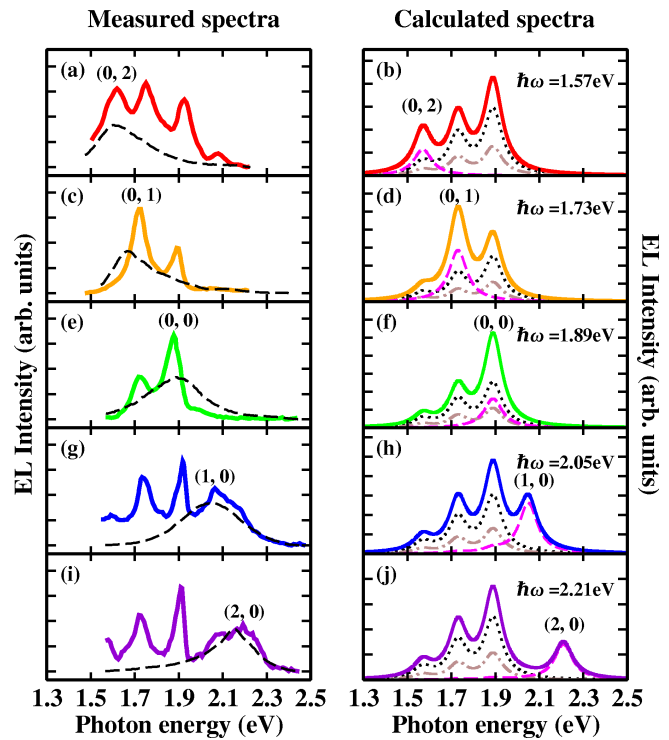}
\caption{\label{fig:fig3} EL spectra modified by the resonant plasmonic excitation.
(Left) Measured EL spectra of TPP molecules together with the corresponding NCP modes
(shown as black dashed line). Adapted by permission from Macmillan Publishers Ltd:
[Nature Photonics] (Ref.\onlinecite{dong_natphotonics_2009}), copyright (2009).
(Right) Calculated EL spectra together with the plasmon-assist molecular emission
(dashed lines), electron tunneling induced emission (dotted lines) and the spontaneous
emission (dash-dotted lines). The plasmon resonant energy is set to 1.57, 1.73, 1.89, 2.05
and 2.21eV, respectively, with the field strength of $1.0\times 10^{6} \mathrm{V/m}$
for the cases (b), (d) and (f), and $5.0\times 10^{6} \mathrm{V/m}$ for (h) and (j).
All the spectra are calculated at +2.5 V and 80 K.}
\end{center}
\end{figure}

It can be seen that the electron tunneling induced electroluminescence strictly follows
the Kasha's rule, i.e. only the emissions from the lowest vibrational level of the
excited state are possible. In the latest experiment, the EL from higher vibrational
levels were observed, which suggests that a new mechanism must be
involved\cite{dong_natphotonics_2009}. One of the possible candidates is certainly the
plasmonic excitation. In order to be consistent with the experiment, we have considered
five cases where the plasmon is resonant with the transition of (0,2), (0,1),
(0,0), (1,0) and (2,0), respectively, as shown in Fig.~{\ref{fig:fig3}} (left).
In our simulations, these transitions correspond to the excitation energy of
1.57, 1.73, 1.89, 2.05 and 2.21 eV, respectively. It is found from the simulations
that the effect of the plasmonic excitation becomes visible for all the five cases
when the field strength of the plasmon is larger than $10^{6}\mathrm{V/m}$.

The simulated EL spectra with the inclusion of both electron tunneling and plasmonic
excitation are given in Fig.~{\ref{fig:fig3}} to directly compare with its
experimental counterpart. One can immediately notice that the spectral feature is drastically
changed by the inclusion of the plasmonic resonant excitations. In other words, all the
non-FC features in Fig.~{\ref{fig:fig3}}, such as new peaks, (1,0) in (h) and (2,0) in (j),
the enhanced peaks, (0,2) in (b), (0,1) in (d) and (0,0) in (f), are contributed
from the plasmonic excitation.  We have also plotted out the contributions from three
individual emission processes in Fig.~{\ref{fig:fig3}} (right). It can be found that
the electron tunneling induced emission and the spontaneous
fluorescence always follow the Kasha's rule and give the conventional double-peak {\it{Q}} profile.
The plasmon-assist molecular emission is a stimulated process directly from the
resonantly excited vibrational levels. The position of such a stimulated emission is
controlled by the plasmonic resonant energy. It is noticed that the agreement between the
theory and the experiment could be largely improved once the plasmon profile is fully
considered in the simulations.

In summary, we have presented a generalized density matrix formalism to successfully
describe the electroluminescence from molecules near a metal surface in a STM.
Both the electron tunneling and local surface plasmon induced excitation and emission
are treated on an equal footing. Model calculations for porphyrin derivatives have
reproduced corresponding experimental spectra and nicely explained the observed unusual
large variation of emission spectral profiles in recent experiments.

Acknowledgement: The work is supported by the National Natural Science Foundation of China (20925311),
the Major State Basic Research Development Programs (2010CB923300), and the Swedish Research Council.

\end{document}